\documentclass[aps,onecolumn,groupedaddress]{revtex4}

\usepackage{graphics}
\usepackage{graphicx}

\def\fun#1#2{\lower3.6pt\vbox{\baselineskip0pt\lineskip.9pt
\ialign{$\mathsurround=0pt#1\hfil##\hfil$\crcr#2\crcr\sim\crcr}}}

\begin{document}
\begin{flushright}
Preprint JLAB-PHY-05-379
\end{flushright}

\title{Single--Spin Asymmetries in the Bethe--Heitler Process
$e^- + p \rightarrow e^- + \gamma + p$ \\
from QED Radiative Corrections}

\date{\today}
\author{Andrei V. Afanasev$^{a)}$, M.I. Konchatnij$^{b)}$ and N.P. Merenkov$^{b)}$}
\affiliation{
$^{(a)}$Thomas Jefferson National Accelerator Facility, Newport News, VA 23606, USA\\
$^{(b)}$ NSC Kharkov Institute of Physics and Technology,
Kharkov 61108, Ukraine}

\begin{abstract}
We derived analytic formulae for the polarization single--spin 
asymmetries (SSA) in the Bethe--Heitler  process $e^- + p
\rightarrow e^- + \gamma + p$. The asymmetries arise due to
one-loop QED radiative corrections to the leptonic part of the
interaction and present a systematic correction for the studies of
virtual Compton Scattering on a proton through interference with
the Bethe-Heitler amplitude.  Considered are SSA with either longitudinally
polarized electron beam or a polarized proton target.
The computed effect appears to be small, not exceeding 0.1 per cent for 
kinematics of current virtual Compton scattering experiments. 
\end{abstract}
\maketitle


\section{Introduction}
Experiments on virtual Compton scattering (VCS) and deeply-virtual Compton scattering (DVCS) 
are an important
part of nucleon structure studies
at major electron-scattering
laboratories, and at Jefferson Lab in particular \cite{JLab12GeV}.
The reason for such a high interest is that VCS allows to access
3-dimensional parton distributions of a nucleon and has a potential of 
resolving the role of parton orbital angular momentum in the nucleon spin problem,
see the original papers \cite{GPD} and recent reviews \cite{Diehl:2003ny, Belitsky:2005qn}.

In experimental observations, the VCS amplitude of electroproduction
of real photons competes with a large and often dominant Bethe-Heitler (BH)
amplitude, in which the real photon is emitted by leptons in the scattering 
process. In the leading order in electromagnetic interactions,
single-spin asymmetries (SSA) in electroproduction of real photons
are caused only by the VCS amplitude, including its interference with 
BH process. BH amplitude alone does not
lead to SSA, unless higher-order electromagnetic corrections are included.
Here we present a calculation of SSA coming from such corrections.

The fact that QED loop radiative corrections can induce the beam
SSA in $\gamma - p$ and $e^- - p$
collisions with production of a $e^+e^-$-pair or radiation of a
photon is known for a long time. A spin-momentum correlation in
differential cross sections for these processes were first
considered by Olsen and Maximon in 1959 \cite{OM}, and somewhat
later in other publications \cite{Oth}. It was shown that a large
azimuthal asymmetry originates from interference of the first and
second (two--photon exchange) Born approximation. These studies
addressed primarily the case of low energies, such that
$\omega/M<<1$, where $\omega$ is the energy of the incoming
polarized particle and $M$ is the target mass.

More recently it was pointed out, for example, in Ref.\cite{AKMP} that
SSA can be induced also by a pure loop correction to the lepton
part of the interaction. For the related processes of radiative
M\"oller scattering and electron-positron pair production on an electron target,
the calculation of Ref.\cite{AKMP} yields substantial SSA, reaching
tens of per cent in selected kinematics. Asymmetries of this magnitude could
present a significant systematic correction to VCS studies, and this fact
partly motivated our present calculation spanning a broad kinematic range. 
In Ref.\cite{Mark} the corresponding
effect is included in the full radiative correction to beam
SSA due to interference between BH and (the absorptive
part of) nucleon VCS amplitudes. In
Ref.\cite{Mark}  the full correction to the beam SSA was computed
numerically, but up to now there is no published analytic
expressions  for the beam SSA caused by the loop correction to the
leptonic tensor, that is simple enough to include in Monte Carlo
generators for analysis of VCS experiments.
The present paper aims at closing this gap. In addition, using
an unpolarized leptonic Compton tensor originally derived in 
Ref.\cite{KMF}, we obtain
new results for SSA in the BH process with a polarized proton target
arising from QED loop corrections. 

The QED effects considered in this paper are referred to as model-independent,
since they do not require additional knowledge of the nucleon
structure. They represent systematic corrections to SSA
measurements such as \cite{CLAS1, HERMES, CLAS2, HallA, d'Hose:2002us} in the interference
region between BH and VCS amplitudes of
electroproduction of real photons. 

\section{General Formalism and Beam Single-Spin Asymmetry}

The contribution to the beam SSA for the BH process,
\begin{equation}\label{1}
e^-(k_1) +P(p_1)\rightarrow e^-(k_2) + \gamma(k) + P(p_2),
\end{equation}
induced by one--loop corrections to the leptonic part of interaction in the case of
longitudinally polarized incoming electrons can be written in terms
of contraction of the leptonic and hadronic tensors
\begin{equation}\label{2}
A^b=\frac{\alpha}{4\pi}\frac{\Re e [P^{(1)}_{\mu\nu}H_{\mu\nu}]}{B_{\mu\nu}H_{\mu\nu}}\  ,
\end{equation}
where the symbol $\Re e$ denotes the real part and
 $B_{\mu\nu}$ is an unpolarized leptonic tensor for large-angle photon emission in the
process (1). We define it as
\begin{eqnarray}
B_{\mu\nu}&=&\frac{1}{4}Tr(\hat{k}_2+m)O_{\lambda\mu}(\hat{k}_1+m)O_{\nu\lambda}\ , \\ \nonumber
O_{\lambda\mu}&=&\gamma_{\lambda}\frac{\hat{k}_2+\hat{k}+m}{s}\gamma_{\mu} +
\gamma_{\mu}\frac{\hat{k}_1-\hat{k}+m}{t}\gamma_{\lambda}\ , s=2kk_2, \ t=-2kk_1\ , \nonumber
\end{eqnarray}
 where $m$ is the electron mass. Note that
both Mandelstam invariants ($s$ and $t$) in experiments that study
proton VCS \cite{CLAS1,HERMES, HallA,CLAS2, d'Hose:2002us} are large, therefore in our
calculations we can omit the lepton  mass in the quantity
$B_{\mu\nu}$ and write the latter in the form
\begin{equation}\label{3}
B_{\mu\nu}=\frac{(s+u)^2+(t+u)^2}{st}\tilde{g}_{\mu\nu} +
\frac{4\Delta^2}{st}(\tilde{k}_{1\mu}\tilde{k}_{1\nu} + \tilde{k}_{2\mu}\tilde{k}_{2\nu})\ , \\
 u= -2k_1k_2\ , \ \Delta=k_2-k_1+k=p_1-p_2,\ \Delta^2=s+t+u,
\end{equation}
where the tilde notation for 4-vectors denotes the gauge-invariant substitution,
$\tilde a_{\mu}= a_{\mu}-\Delta_{\mu}\frac{a\Delta}{\Delta^2}$.

\begin{figure}[t]
\includegraphics[width=0.9\textwidth]{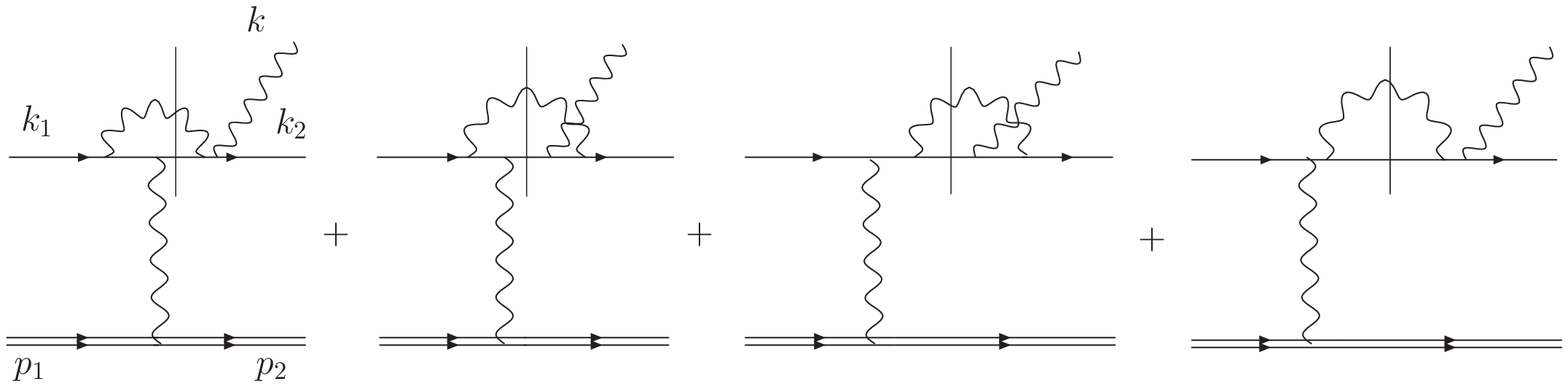}
\parbox[t]{0.85\textwidth}
{\caption{\small{One-loop corrected diagrams that induce
single-spin asymmetries in the BH process (1). 
Only radiation from the outgoing electron contributes.
Vertical solid lines indicate the unitarity cuts.
}
} 
\label{diagrams} }
\end{figure}

For the hadronic tensor, we use its Born expression
$$H_{\mu\nu}=\frac{1}{4}Tr(\hat{p}_2+M)\Gamma_{\mu}(\hat{p}_1+M)(1-\gamma_5
\hat{S}) \Gamma_{\nu}\ ,\ \Gamma_{\mu}=(F_1+F_2)\gamma_{\mu}
-\frac{p_{1\mu}+p_{2\mu}}{2M}F_2\ , $$ where $F_{1,2}\equiv F_{1,2}(\Delta^2)$ are
the Dirac and Pauli proton form factors, respectively, and $S$ is
a 4--vector of proton polarization. The expressions for spin--independent
and spin--dependent parts of the hadronic tensor are
\begin{eqnarray}
&&H^{(un)}_{\mu\nu}=\frac{\Delta^2}{2}\big(F_1+F_2)^2\tilde{g}_{\mu\nu}+2\Big(F_1^2-
\frac{\Delta^2}{4M^2}F_2^2\Big)\tilde{p}_{1\mu}\tilde{p}_{1\nu}\ , \\
&&H^{S}_{\mu\nu}=-iM(F_1+F_2)(\mu\nu
\Delta\rho)\Big[\Big(F_1+\frac{\Delta^2}{4M^2}F_2\Big)S_{\rho}
+\frac{F_2(Sp_2)}{2M^2}p_{1\rho}\Big] \ ,
\end{eqnarray} where $(\mu\nu
\Delta\rho)\equiv \epsilon_{\mu\nu\lambda\rho}\Delta_{\lambda}$ and the sign convention for
the Levi-Civita tensor is $\epsilon_{0123}=1.$ Only the spin-independent hadronic
tensor contributes to the beam SSA,
therefore contraction of the tensors reads
\begin{eqnarray}\label{4}
H^{(un)}_{\mu\nu}B_{\mu\nu}
&=&\frac{(s+u)^2+(u+t)^2}{st}\Big[\Delta^2(F_1+F_2)^2 +
2M^2\Big(F_1^2-\frac{\Delta^2}{4M^2}F_2^2\Big)\Big]+\nonumber \\
&&\frac{2\Delta^2}{st}\Big(F_1^2-\frac{\Delta^2}{4M^2}F_2^2\Big)\big[V^2+X^2+V(u+t
)-X(u+s)\big]\ , 
\end{eqnarray}
where $V=2k_1p_1, \ X=2k_2p_1 $.

Let us now consider the leptonic tensor. In general, its spin-dependent part
for the case of the beam longitudinal polarization can be written as
\begin{equation}\label{5}
P_{\mu\nu}=P^{(0)}_{\mu\nu}+\frac{\alpha}{4\pi}P^{(1)}_{\mu\nu}\ ,
\end{equation}
where the Born value $P^{(0)}_{\mu\nu}$ is pure imaginary and antisymmetric and therefore
it does not contribute in the beam SSA, whereas the one-loop correction to it
$P^{(1)}_{\mu\nu}$, together with the imaginary antisymmetric part, contains also the the real and
symmetric part. The latter is caused by interference between the Born-level
BH amplitude and the diagrams in Fig.\ref{diagrams} with an additional photon loop and 
an external photon coupling to the outgoing electron.
These diagrams produce a nonzero imaginary (absorptive) 
part of the amplitude in the physical region of the process (1) due to two-particle electron-photon
intermediate states as shown in Fig.\ref{diagrams} by the unitarity cuts.

Calculation of the one-loop diagrams of Fig.\ref{diagrams} contributing to the radiative leptonic tensor
can be done by standard QED techniques. 
Details of the calculation can be found in Ref.\cite{KMF} for unpolarized electrons and in 
Ref.\cite{AAK} for longitudinally polarized electrons.
In calculations of SSA, we are only concerned with the absorptive parts
of the electron virtual Compton amplitude that enters the radiative leptonic tensor.
It should be noted that calculation of SSA is infra-red safe, {\it i.e.}, 
infra-red singularities explicitly cancel at the loop level for this observable.
It is an important consistency check of the calculation, since ultra-soft photon radiation
that normally cancels such singularities has zero absorptive part and therefore
cannot assist with infra-red divergence cancellation in the considered observable.

The contribution to $P^{(1)}_{\mu\nu}$
we are interested in can be derived by standard loop computation
techniques of QED. Neglecting terms that are explicitly anti-symmetric in indices $\mu \leftrightarrow \nu$,
we arrive at the following expression ($c.f.$ Eq.(36) of Ref.\cite{AAK}),
\begin{equation}\label{6}
P^{(1)}_{\mu\nu}=i(k_1 k_2 \Delta \nu)[B_1 \tilde{k}_{1\mu}+B_2 \tilde{k}_{2\mu}]-i(k_1
k_2 \Delta \mu)[ B^*_1 \tilde{k}_{1\nu}+B^*_2 \tilde{k}_{2\nu}]\ ,
\end{equation}
where the quantity $B_1$ has the form
\begin{eqnarray}\label{7}
B_1&=&\frac{2}{st}\Big[\frac{8u}{a}\big(1-\frac{\Delta^2}{a}L_{\Delta u}\big)+\frac{6t}{b}L_{\Delta t}+
\frac{2(u^2-2s^2-su)}{cu}L_{s\Delta}+\\ \nonumber
&&\frac{2b}{c}\big(1+\frac{s}{c}L_{s\Delta}\big)+\frac{2}{s}(2c-s)L_{tu} +\big(-2-
\frac{4c^2}{st}-\frac{12b}{t}-\frac{4s^2}{ut}\big)L_{\Delta u}+\frac{4b^2}{ut}L_{su}+\\ \nonumber
&&\big(-2 + \frac{2uc}{s^2}-\frac{2t}{s}\big)G +\big(\frac{2b}{t}+\frac{2b^2}{t^2}\big)
\widetilde{G} +6\Big]\ , 
\end{eqnarray}
where, except for the 4-vector $\Delta$, we used the same notation as in \cite{AAK} (see also \cite{KMF}),
\begin{eqnarray} 
(k_1 k_2 \Delta \sigma)&=&\epsilon_{\alpha\beta\lambda\sigma}k_{1\alpha} k_{2\beta}
 \Delta_{\lambda}\ , \ a =s+t, \ b= s+u, \ c= t+u\ ,\\ \nonumber
G&=&L_{\Delta u}(L_\Delta+L_u-2L_t)-\frac{\pi^2}{3}-2Li_2\big(1-\frac{\Delta^2}{u}\big)+2Li_2\big(
1-\frac{t}{\Delta^2}\big)\ , \\ \nonumber 
L_{xy} &=& L_x -L_y\ , \ \
L_\Delta=\ln\frac{-\Delta^2}{m^2}\ , \ L_u=\ln\frac{-u}{m^2}\ , \
L_t=\ln\frac{-t}{m^2}\ , \ L_s=\ln\frac{-s}{m^2}\ . \nonumber 
\end{eqnarray}
The quantities
$B_2$ and $\widetilde{G}$ can be derived from $B_1$ and $G$ by
substitution $$B_2=-B_1(s\leftrightarrow t), \ \
\widetilde{G}=G(s\leftrightarrow t)\ .$$

Imaginary parts of $B_1$ and $B_2$ which induce the real symmetric part of
$P^{(1)}_{\mu\nu}$ can arise from the terms containing $L_s$ and $\widetilde{G},$ and the terms contributing
to the imaginary part of $\widetilde{G}$ are $L_s$ and $Li_2(1-s/\Delta^2)$ because of the condition
$1-s/\Delta^2 > 1.$
From the form of proparators in the Feynman diagrams of Fig.\ref{diagrams} it follows that to obtain
the imaginary part of $B_1$ and $B_2$, one has to add a small negative imaginary part to the
electron mass. It leads to
\begin{equation}\label{8}
L_s=\ln\frac{s}{m^2} -i\pi, \ \ Li_2\big(1-\frac{s}{\Delta^2}\big) =
-\int\limits_0^{1-s/\Delta^2}
\frac{dx\ln{x}}{1-x} -\ln\frac{u+t}{\Delta^2}L_{s\Delta}\ ,
\end{equation}
and it means that
$$\Im m\widetilde{G}=2\pi\ln\frac{u+t}{u}\ ,$$
where the symbol $\Im m$ stands for the imaginary part.
By combining the previous results, we arrive at
\begin{eqnarray}\label{9}
\Im m B_1 &=&-\frac{2\pi}{st}\bar{B_1}\ , \ \ \Im m B_2 =-\frac{2\pi}{st}\bar{B_2}\ ,\nonumber \\ 
\bar{B}_1&=&\frac{2 \Delta^2 (\Delta^2-t)}{t}\bigl[\frac{3t+2u}{(t+u)^2}-\frac{2}{t}\ln(1+\frac{t}{u})\bigr],\\
\bar{B}_2&=&\frac{2\Delta^2}{t}\bigl[\frac{t-2u}{t+u}-2(1-\frac{u}{t})\ln(1+\frac{t}{u})\bigr] \nonumber
\end{eqnarray}
Note that the terms containing anomalous poles at $t\rightarrow 0$ are cancelled explicitly
in Eqs. (\ref{9}). Moreover, both $\bar{B}_1$ and $\bar{B}_2$
are proportional to $t$ if $t\to 0$.
Contraction of the tensors in the numerator of Eq.(\ref{2}) can be written in the following 
form that appears as a factor in the spin-dependent numerator for beam SSA:
\begin{equation}\label{11}
P^{(1)}_{\mu\nu}H_{\mu\nu} =
\frac{2\pi(k_1k_2\Delta p_1)}{st}\Big(F_1^2- \frac{\Delta^2}{4M^2}F_2^2\Big)
[(2V-s+\Delta^2)\bar{B}_1+(2X-s-u)\bar{B}_2]\ .
\end{equation}
An important observation is that the expressions (\ref{9}) contain an overall
factor of $\Delta^2$, that is a squared 4-momentum transferred to the proton target.
Since experiments on DVCS select small values of $\Delta^2$, it directly
implies additional suppression of beam SSA in this region.


Equations (\ref{2}),(\ref{4}), (\ref{9}-\ref{11}) determine the beam SSA
for the BH process (\ref{1}). Note that the expressions for SSA do not
contain large logarithms involving the lepton mass. 

Let us now introduce kinematic
invariants used to define experimentally measured asymmetries in VCS.
The following set of kinematic variables \cite{Bal} is accepted for experimental 
analyses: Azimuthal angle $\Phi$ between planes $({\bf q}_1,
{\bf k}_1)$ and $({\bf q}_1, {\bf p}_2)$ in laboratory system
(with axes $OZ$ along direction $-{\bf q}_1= {\bf k}_2-{\bf
k}_1),$ and invariant variables
\begin{equation}\label{12}
x=\frac{-q_1^2}{2p_1q_1}\ , \ \ y=\frac{2p_1q_1}{V}\ , \ \
\ \ q_1=k_1-k_2,\ q_1^2\equiv u.
\end{equation}

The advantage of these variables is that one may use both the
laboratory system and c.m.s. of subprocess $\gamma(q_1)+P(p_1)
\rightarrow \gamma(k)+P(p_2)$ to investigate the azimuth
correlation. The reason is that c.m.s. can be reached from
laboratory system by a boost along the direction of ${\bf q}_1,$ and
such transformation does not change the azimuthal angle $\Phi .$
Therefore, we can use the simple expressions for the particle
energies and their angles in c.m.s. 
\begin{eqnarray}
p_1&=&(E_1,0,0,|{\bf p}_1|), \
p_2=(E_2,|{\bf p}_2|\sin{\theta_N}\cos{\Phi},|{\bf
p}_2|\sin{\theta_N}\sin{\Phi},|{\bf p}_2|\cos{\theta_N}),\\ \nonumber 
k_1&=&\varepsilon_1(1,\sin{\theta_1},0,\cos{\theta_1}), \
k_2=\varepsilon_2(1,\sin{\theta_2},0,\cos{\theta_2}), \\
k&=&(k_0,-{\bf p}_2), \ \ q_1=(\varepsilon_1-\varepsilon_2, -{\bf
p}_1) \nonumber 
\end{eqnarray}
in terms of variables (\ref{12}) to form all the
invariants that enter into beam and target asymmetries. They
read
\begin{eqnarray}\label{13}
\varepsilon_1&=&\frac{V(1-xy)}{2\sqrt{\xi V}}\ , \
\varepsilon_2=\frac{V(1-y+xy)}{2\sqrt{\xi V}}\ , 
k_0=\frac{V(y-xy)}{2\sqrt{\xi V}}\ ,\nonumber \\
E_1&=&\frac{V(y+2\tau)}{2\sqrt{\xi V}}\ ,\ 
E_2=\frac{V(y-xy+2\tau)}{2\sqrt{\xi V}}\ ,\nonumber \\
\cos{\theta_1}&=&-\frac{y(1-2x+xy+2x\tau)}{\lambda(1-xy)}\ , \
\cos{\theta_2}=-\frac{y(1-2x-y+xy-2x\tau)}{\lambda(1-y+xy)}\ ,\\ \nonumber
\sin{\theta_1}&=&\frac{2\sqrt{\eta\xi xy}}{\lambda(1-xy)}\ , \ \
\sin{\theta_2}=\frac{2\sqrt{\eta\xi xy}}{\lambda(1-y+xy)}\ ,\\ \nonumber
\cos{\theta_N}&=&\frac{y(y-xy+2x\tau)-2\xi\rho}{\lambda(y-xy)}\ , \
\
\sin{\theta_N}=\frac{2\xi\sqrt{\rho_+-\rho)(\rho-\rho_-)}}{\lambda(y-xy)}\ \nonumber
,
\end{eqnarray}
where we used the following brief notation 
\begin{eqnarray}
\lambda^2&=&y^2+4xy\tau, \ \ \eta=1-y-xy\tau, \ \ \xi=y-xy+\tau, \ \
\tau=\frac{M^2}{V}\ ,\\ \nonumber
 \rho&=&\frac{-\Delta^2}{V}, \ \
\rho_{\pm}=\frac{y}{2\xi}[(1-x)(y\pm \lambda)+2x\tau].
\end{eqnarray} 
Here we introduced a dimensionless variable $\rho$ with
the quantities $\rho_{\pm}$ having a meaning of the minimum
$(\rho_-)$ and maximum $(\rho_+)$ value of $\rho$ at fixed $x, \ y$
and $V.$ By using relations (\ref{13}) it is straightforward to obtain
\begin{eqnarray}\label{14}
& &s=\frac{Vy}{\lambda^2}
[2K\cos{\Phi}+xy(1+2x\tau)+\rho(1-y-2x+xy-2x\tau)]\ , \\ \nonumber
& &t=-\frac{Vy}{\lambda^2}
[2K\cos{\Phi}+xy(1-y-2x\tau)+\rho(1-2x+xy+2x\tau)]\ , \\
& &(k_1k_2\Delta p_1)=-\frac{V^2Ky}{2\lambda}\sin{\Phi}\ , \ \
K^2=\frac{\eta\xi x(\rho_+-\rho)(\rho-\rho_-)}{y}\ .\nonumber
\end{eqnarray}

Our estimation of the beam SSA is demonstrated in Fig.2 for the
conditions of Jefferson Lab and HERMES experiments on beam SSA in VCS
at different electron beam energies: $E_b=$ 4.25 GeV \cite{CLAS1},
5.75 GeV \cite{CLAS2} and 27.5 GeV \cite{HERMES} and fixed values of $-q_1^2$= 1.25, 1.08 and 2.6 GeV$^2$, respectively.
It can be seen from Fig.2 that the asymmetry is rather small, not exceeding
0.1 per cent in the kinematics of the considered experiments, even for a rather broad
range  of variable $\rho$. The calculated effect
is smaller for the kinematics of planned SSA measurements in DVCS by COMPASS
collaboration at CERN \cite{d'Hose:2002us}, since it covers smaller values of $\Delta^2$.
The effect for kinematics \cite{CLAS2} and \cite{HallA} are similar in magnitude.

\begin{figure}[h]\label{bSSA}
\includegraphics[width=0.45\textwidth]{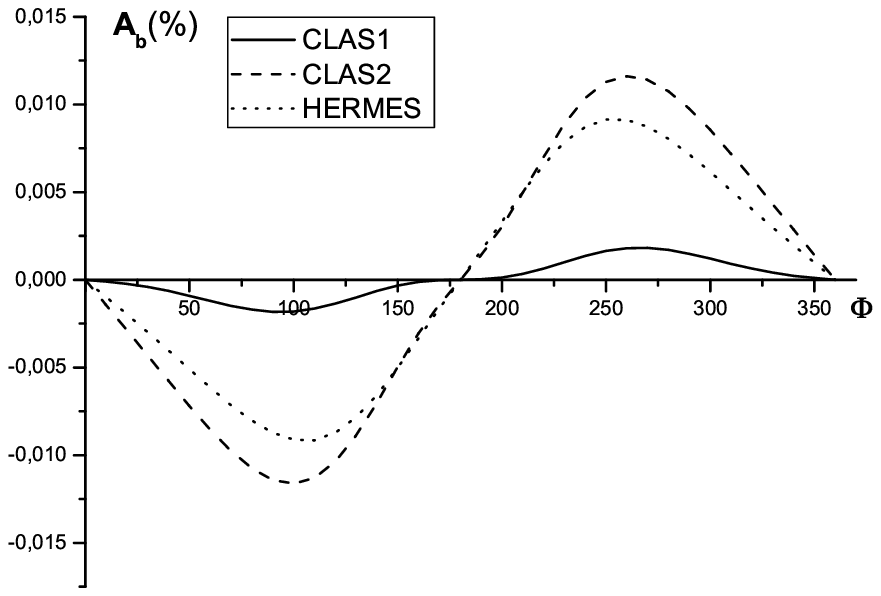}
\includegraphics[width=0.45\textwidth]{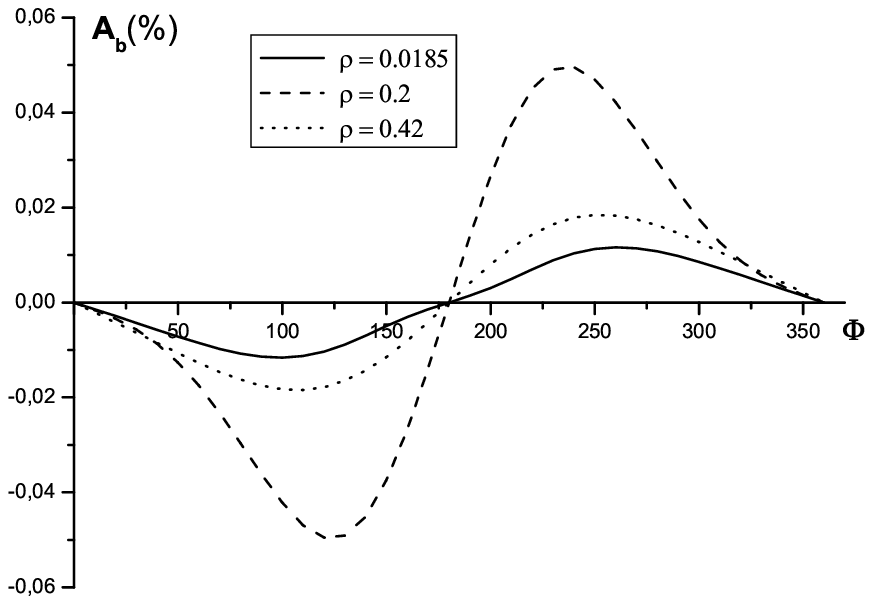}
\parbox[t]{0.85\textwidth}
{\caption{\small{Dependence of the beam SSA (in per cent) on
azimuth angle (in degrees) for three different experimental
conditions (left panel): CLAS1 \cite{CLAS1} corresponds to
$x=0.19, \ y=0.825, \ \rho=0.024,$ for CLAS2 \cite{CLAS2} $x=0.18,
\ y=0.5, \ \rho=0.0185,$ and for HERMES \cite{HERMES} $x=0.11, \
y=0.458, \ \rho=0.005.$ Curves for different values of $\rho$ on
the right panel correspond to CLAS2 conditions \cite{CLAS2}.}}}
\end{figure}

\begin{figure}[!t]
\includegraphics[width=0.4\textwidth]{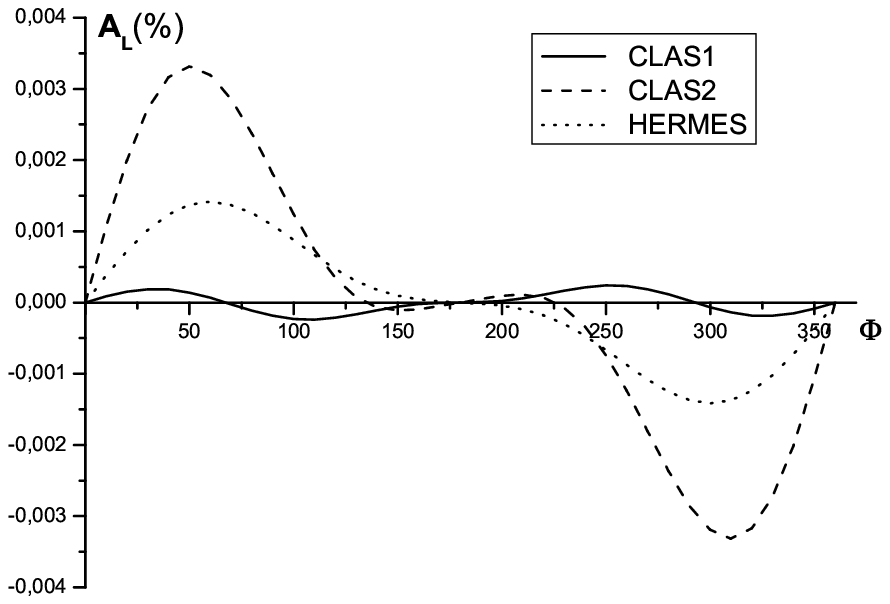}
\includegraphics[width=0.4\textwidth]{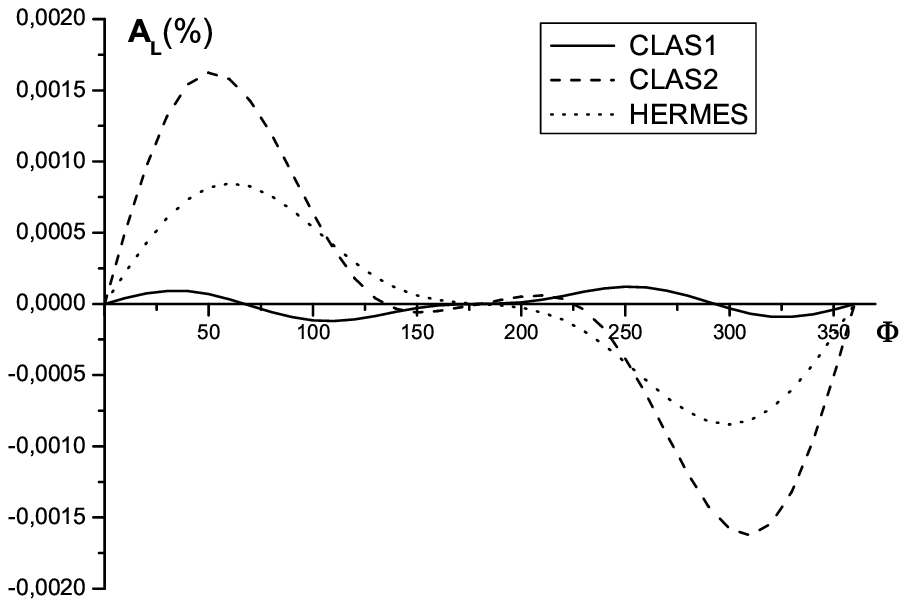}
\includegraphics[width=0.4\textwidth]{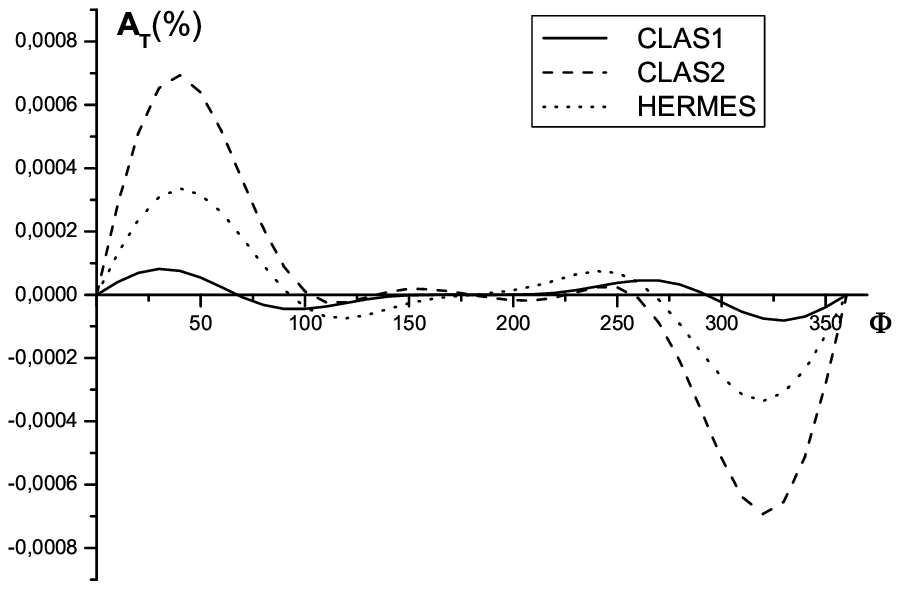}
\includegraphics[width=0.4\textwidth]{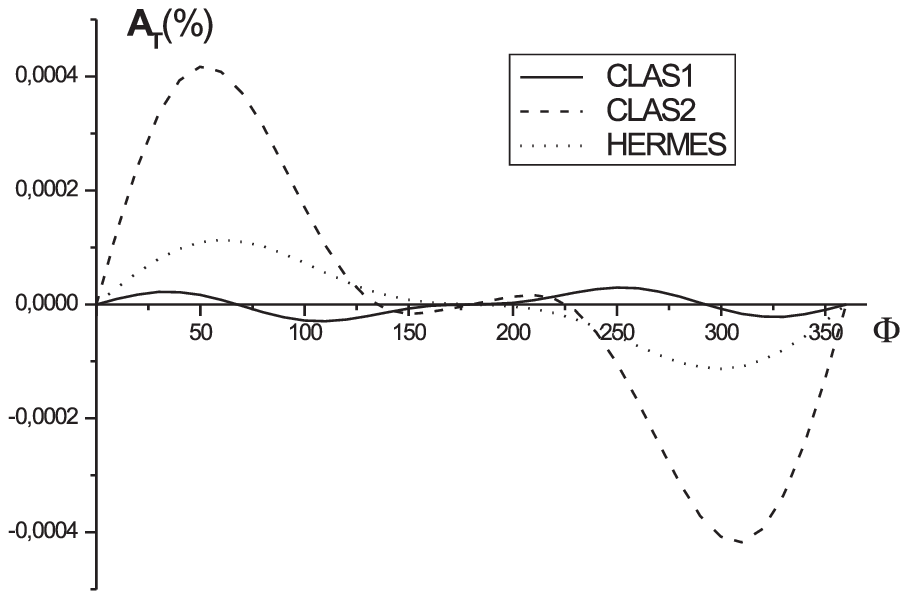}
\includegraphics[width=0.4\textwidth]{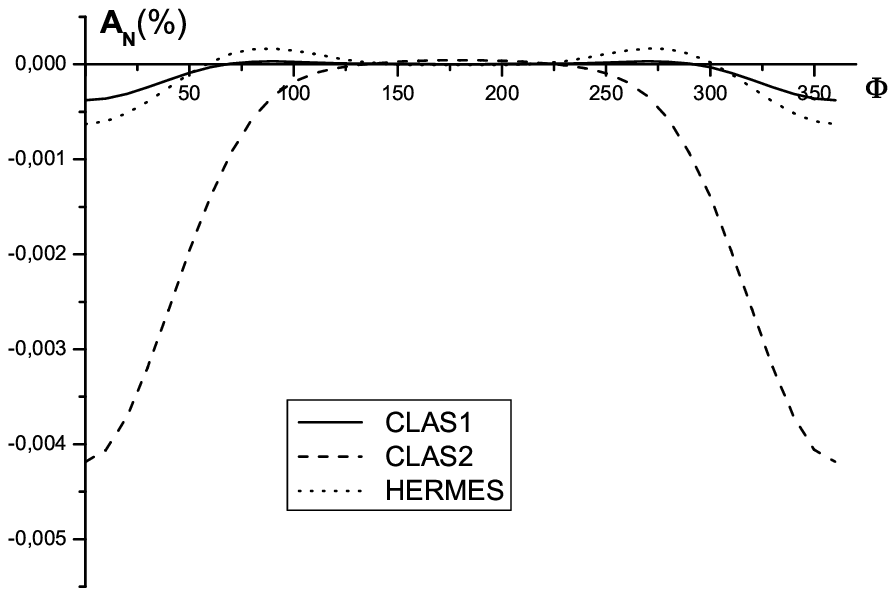}
\includegraphics[width=0.4\textwidth]{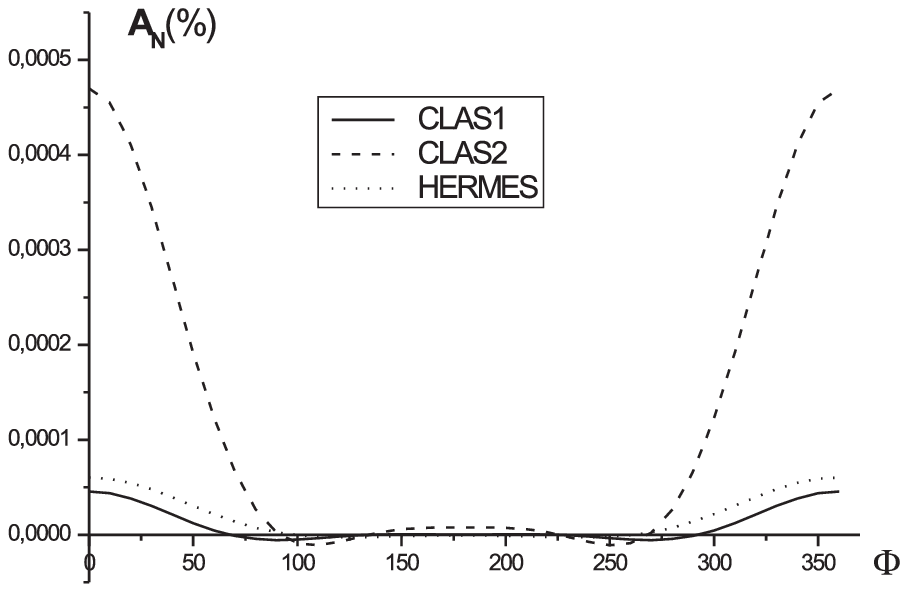}
\parbox[t]{0.85\textwidth}
{\caption{\small{The target SSA as a function of azimuthal angle $\Phi$:
For left panel, the polarization 4-vectors are defined by
Eqs. (\ref{20}) and the right panel corresponds to Eqs.(\ref{24}). }}
\label{targetSSA}}
\end{figure}

\section{Target single--spin asymmetries}

Let us now consider the target SSA in
the process (\ref{1}) caused by one--loop corrections to the unpolarized
part of the leptonic tensor. In contrast with the beam single--spin
correlation, where the effect is due to the real symmetrical part
of the spin--dependent leptonic tensor, this time it is
related with the imaginary antisymmetric part of the
spin-independent leptonic tensor. In this case
\begin{equation}\label{15}
A^p=\frac{\Re e [H^S_{\mu\nu}K^{(1)}_{\mu\nu}]}{B_{\mu\nu}H^{(un)}_{\mu\nu}}\ ,
\end{equation}
where the effect is induced by the imaginary and antisymmetric
part of the spin--independent tensor $K^{(1)}_{\mu\nu}$ that can
be computed directly ($c.f.$ \cite{KMF}). Keeping only antisymmetric terms in
$\mu \leftrightarrow \nu$, the result can be written as
\begin{equation}\label{16}
K^{(1)}_{\mu\nu}=\frac{i\alpha}{4\pi}\Im m[(T_{12}-T_{21})(\tilde{k}_{1\mu}\tilde{k}_{2\nu}-
\tilde{k}_{1\nu}\tilde{k}_{2\mu})]\ ,
\end{equation}
where, neglecting terms proportional to $m^2$, we have
\begin{eqnarray}\label{17}
T_{12}&=&\frac{2}{st}\Big[\frac{\Delta^2}{s^2}c(u-s)G+\frac{\Delta^2}{t^2}(u\Delta^2-st)\widetilde{G}-
2\Delta^2\big(\frac{u\Delta^2}{st}+\frac{2u-s+t}{a}L_{\Delta u}
\big)+8u+3t-s+\frac{2us}{c}- \nonumber \\
&&\frac{4(u^2-cs)(\Delta^2L_{\Delta u}-a)}{a^2}+\frac{\Delta^2(2c+t)(st-u\Delta^2)L_{\Delta s}}{c^2t}-
\frac{\Delta^2c(2u-s)L_{\Delta t}}{bs}\Big] \ ,
\end{eqnarray}
and $T_{21}$ is derived from $T_{12}$ by substitution
$T_{21}=T_{12}(t\leftrightarrow s).$

After extracting absorptive parts from $L_s$ and $\widetilde{G}$ we
obtain the following expression for antisymmetric imaginary
part of the leptonic tensor (\ref{16}):
\begin{equation}\label{18}
K^{(1)}_{\mu\nu}=\frac{i\alpha}{2}(\tilde{k}_{1\mu}\tilde{k}_{2\nu}-
\tilde{k}_{1\nu}\tilde{k}_{2\mu})T\ , \ \
T=\frac{\Delta^2}{st}\Big[\frac{2st}{c^2}
+4u\Big(\frac{1}{t}\ln\frac{u+t}{u}-\frac{1}{c}\Big)\Big]\ .
\end{equation}
We note an overall factor of $\Delta^2$ from Eq.(\ref{18}) that
leads to additional suppression of target SSA in DVCS kinematics.
 
Contraction of the antisymmetric tensors in
expression (\ref{15}) for target SSA reads
\begin{equation}\label{19}
H^S_{\mu\nu}K^{(1)}_{\mu\nu}= \alpha M(F_1+F_2)TG_s \ ,
\end{equation}
where the quantity $G_s$ depends on the target--proton polarization
4--vector $S$ and has the form
\begin{equation}\label{TSSA}
G_s=\Big[\Big(F_1+\frac{\Delta^2}{4M^2}F_2\Big)(k_1k_2\Delta S)+\frac{F_2}{2M^2}(p_2S)
(k_1k_2\Delta p_1)\Big] \ .
\end{equation}

It follows from Eq. (\ref{TSSA}) that in general the one--loop correction
to the leptonic tensor in radiative process (1) generates the
target SSA due to all three possible orientations of the target polarization.
Here we consider two possible conventions for defining
directions of target polarization. First, one can
consider the case when the longitudinal polarization (L) in
laboratory system is along the electron beam direction of ${\bf k}_1,$ transverse
polarization (T) lies in the plane $({\bf k}_1,{\bf k}_2)$ and
the normal (N) one is along the normal vector $({\bf k}_1\times{\bf k}_2).$
The corresponding polarization 4--vectors can be expressed via the
4--momenta \cite{Pol1}
\begin{equation}\label{20}
S^L_{1\mu}=\frac{2\tau k_{1\mu}-p_{1\mu}}{\sqrt{V\tau}}\ , \
S^N_{1\mu}=-\frac{2(\mu k_1k_2p_1)}{\sqrt{V^3xy\eta}}\ , \
S^T_{1\mu}=\frac{k_{2\mu}-(1-y-2xy\tau)k_{1\mu}-xyp_{1\mu}}
{\sqrt{Vxy\eta}}\ .
\end{equation}
Then we have
\begin{eqnarray}
G^L_{s1}&=&\frac{(k_1k_2\Delta p_1)}{\sqrt{V\tau}}\Big[-F_1+\frac{F_2}{2}
(\tilde t-xy)\Big]\ , \ \ \tilde t = \frac{t}{V}\ ,\\ \nonumber
G^T_{s1}&=&\frac{(k_1k_2\Delta p_1)}{\sqrt{Vxy\eta}}\Big[-xyF_1+\frac{F_2}{4\tau}
\big(xy(1-y-2xy\tau)+\tilde ty(1+2x\tau)+\rho\big)\Big]\ ,\\
G^N_{s1}&=&-\frac{1}{4}\sqrt{\frac{V^3xy}{\eta}}
\bigg\{\Big(F_1-\frac{\rho}{4\tau}F_2\Big)[xy(1-y)-\rho(1-xy)-(2-
y)\tilde
t] -\frac{4F_2(k_1k_2\Delta p_1)^2}{V^4xy\tau}\bigg\}\ ,\nonumber
\end{eqnarray}
with the same proton form factors $F_{1,2}$ as in Eqs.(5,6).

In another convention, one may choose directions to define the polarization 3--vector 
in the lab system. If the longitudinal direction is taken along
${\bf p}_2,$ the transverse one in the plane $({\bf k}_1, {\bf p}_2)$ and
the normal is along 3--vector ${\bf p}_2\times{\bf k}_1,$ then the corresponding
polarization 4--vector can be written as \cite{Pol2}
\begin{eqnarray}\label{24}
S^L_{2\mu}&=&\frac{-2\tau \Delta_{\mu}-\rho
p_{1\mu}}{\sqrt{V\tau\rho(4\tau +\rho)}}\ , \
S^N_{2\mu}=-\frac{2(\mu p_2k_1p_1)}{\sqrt{V^3\zeta}}\ , 
\zeta=\rho(1-xy+\tilde t)-\tau(xy-\tilde t)^2\ ,\\
S^T_{2\mu}&=&\frac{\rho(4\tau+\rho)k_{1\mu}+(\rho-2\tau(\tilde
t-xy) \Delta_{\mu}-\rho(2-xy+\tilde
t)p_{1\mu}}{\sqrt{V\zeta\rho(4\tau+\rho)}}\ . \nonumber
 \end{eqnarray}
In this case
\begin{eqnarray}\label{25}
G^L_{s2}&=&-\sqrt{\frac{\rho}{V\tau(4\tau+\rho)}}(k_1k_2\Delta p_1)(F_1+F_2)\ ,\nonumber \\
G^T_{s2}&=&-\sqrt{\frac{\rho}{V\zeta(4\tau+\rho)}}(k_1k_2\Delta p_1)(2-xy+\tilde
t)\Big(F_1-\frac{\rho}{4\tau}F_2\Big)\ ,\\ 
G^N_{2s}&=&\frac{1}{4}\sqrt{\frac{V^3}{\zeta}}\big\{\rho[\tilde t(1+xy)
+xy(1-xy)]+y(\tilde t-xy)(\tilde t+x-xy)\Big\}
\Big(F_1-\frac{\rho}{4\tau}F_2\Big)\ .\nonumber
\end{eqnarray}

The target SSA in the BH process (1) is shown in Fig.3. Our
calculations indicate that beam and target SSA
generated by loop corrections to leptonic part of the interaction in
the BH process are small and for the considered experimental
conditions they do not exceed 0.1 per cent. The reason is that
in addition to being multiplied by the fine structure constant $\alpha=\frac{1}{137}$,
they contain additional suppression for the relevant values of kinematic 
invariants. 

\section{Summary}

In conclusion, let us discuss the role of other radiative corrections
in the BH process coming from real-photon radiation. In VCS experiments, the kinematic cuts
are imposed in such a way that the phase space of the (undetected) additional 
photon is restricted to its relatively small values. In this
case, the main contribution to the radiative correction comes from spin-independent
soft photon emission that does not affect polarization observables,
but does change unpolarized cross sections by as much as about 20 per cent
(see, e.g., \cite{Mark} for VCS case and Ref.\cite{Pol2, Afanasev:2001nn} for elastic electron-proton scattering). 
Therefore in such a soft-photon-emission regime,
the loop correction considered here is {\it the only} model-independent radiative correction
to SSA.

 Thus, we demonstrated that systematic corrections to beam and target SSA
 arising from the higher-order QED effects in the BH process are negligible
 compared to the relatively large (tens of per cent) experimentally observed asymmetries \cite{CLAS1,HERMES}
 due to interference between the BH and VCS amplitudes in electroproduction of real photons.
 We confirm that the basic assumption of negligible SSA from the BH process alone holds to
 better than 0.1 per cent accuracy, thereby justifying present interpretation of
 SSA as arising mainly from the BH-VCS interference and VCS mechanisms.

\section*{Acknowledgements}

This work was supported by the US Department of Energy
under contract DE-AC05-84ER40150.

\end{document}